\documentclass[prl,aps,twocolumn,amsmath,amssymb]{revtex4}

\pdfoutput=1
\usepackage{hyperref}
\usepackage{graphicx}
\usepackage[final]{movie15}

\def\eqq#1{Eq.~(\ref{#1})}

\def\f#1{Fig.~\ref{#1}}

\def\c#1{~\cite{#1}}

\def\ed{\epsilon_{\rm d}}
\def\es{\epsilon_{\rm s}}

\def\e{{\rm e}}

\def\beq{\begin{equation}}
\def\eeq{\end{equation}}
\def\bea{\begin{eqnarray}}
\def\eea{\end{eqnarray}}

\def\kt{k_{\rm B}T}

\begin{document}

\title{Self-assembly at a nonequilibrium critical point}

\author{Stephen Whitelam\footnote{\tt{swhitelam@lbl.gov}}$^1$, Lester O. Hedges$^1$, Jeremy D.\ Schmit$^2$}
\affiliation{$^1$ Molecular Foundry, Lawrence Berkeley National Laboratory, 1 Cyclotron Road, Berkeley, CA 94720, USA \\
$^2$ Department of Physics, Kansas State University, Manhattan, KS 66506, USA}
\begin{abstract}
We use analytic theory and computer simulation to study patterns formed during the growth of two-component assemblies in 2D and 3D. We show that these patterns undergo a nonequilibrium phase transition, at a particular growth rate, between mixed and demixed arrangements of component types. This finding suggests that principles of nonequilibrium statistical mechanics can be used to predict the outcome of multicomponent self-assembly, and suggests an experimental route to the self-assembly of multicomponent structures of a qualitatively defined nature.
\end{abstract}
\maketitle

Nature builds its materials using multiple component types. The properties of these materials depend not just on the properties of their components, but on how components are distributed spatially: the exciton transfer properties of a light-harvesting complex, for instance, depend both on the proteins that comprise them and on the way proteins cluster\c{bahatyrova2004native}. Achieving similar mesoscale spatial control with many synthetic self-assembled materials\c{kong2013mapping} is made difficult by the fact that the self-assembly of multicomponent systems generally happens `far' from equilibrium, where we possess few predictive theoretical tools. Although self-assembly is a nonequilibrium process, much of our understanding of it is based upon a physical picture that assumes dynamics to play no role {\em except} to convey a system along the `easiest' pathways on its free energy landscape\c{stranski1933rate}. This `near-equilibrium' or `quasiequilibrium' assumption tends to hold, for instance, for simple one-component systems under mild nonequilibrium conditions\c{wolde1999homogeneous,shen1996bcc,sear2007nucleation,wolde1997epc,lutsko2006ted}. It fails when there exist timescales within a given self-assembly process that exceed the time of the experiment or computer simulation. For one-component systems, long timescales can emerge if bonds between particles are strong, and so break infrequently, which can happen when subjected to `harsh' nonequilibrium conditions (e.g. conditions of deep supercooling). Mistakes of binding made as components associate then fail to anneal as structures grow, and the result is a kinetically trapped structure rather than an object corresponding to a favored position on the free energy landscape\c{hagan2006dynamic,wilber2007reversible,grant2012quantifying,haxton2012design}.

The self-assembly of multicomponent solid structures, however, is {\em also} affected by kinetic traps that emerge even under {\em mild} nonequilibrium conditions. The long timescale responsible for such trapping is the slow interchange of component types within solid structures\c{kremer1978multi,stauffer1976kinetic,PhysRevB.27.7372,schmelzer2004nucleation,schmelzer2000reconciling,scarlett2010computational,kim2008probing,scarlett2011mechanistic,sanz2007evidence,peters2009competing,whitelam2012self}. Consider \f{fig1}(a), which shows in cartoon form a fluid of two types of mutually attractive particle (call them `red' and `blue'), present in equal number, self-assembling into a solid structure. The equilibrium pattern of red and blue within this structure -- shown in the figure as demixed red and blue domains -- is determined only by particles' mutual interactions. But achieving this equilibrium pattern via self-assembly requires that particle types interchange their positions within the assembly  readily enough to `unmix' themselves. Because the mobility of particles within solid structures is usually very low (red arrows on figure), such interchange can fail to happen on experimental timescales. The result is a structure whose component types are distributed in a nonequilibrium manner\c{scarlett2010computational,kim2008probing,scarlett2011mechanistic,sanz2007evidence,peters2009competing,whitelam2012self}. Returning to the picture of evolution on a free energy landscape, one must think of the direction of motion across this landscape as being biased strongly by the underlying microscopic dynamics\c{peters2009competing}.
\begin{figure*}[t!]
\centering
\includegraphics[width=\linewidth]{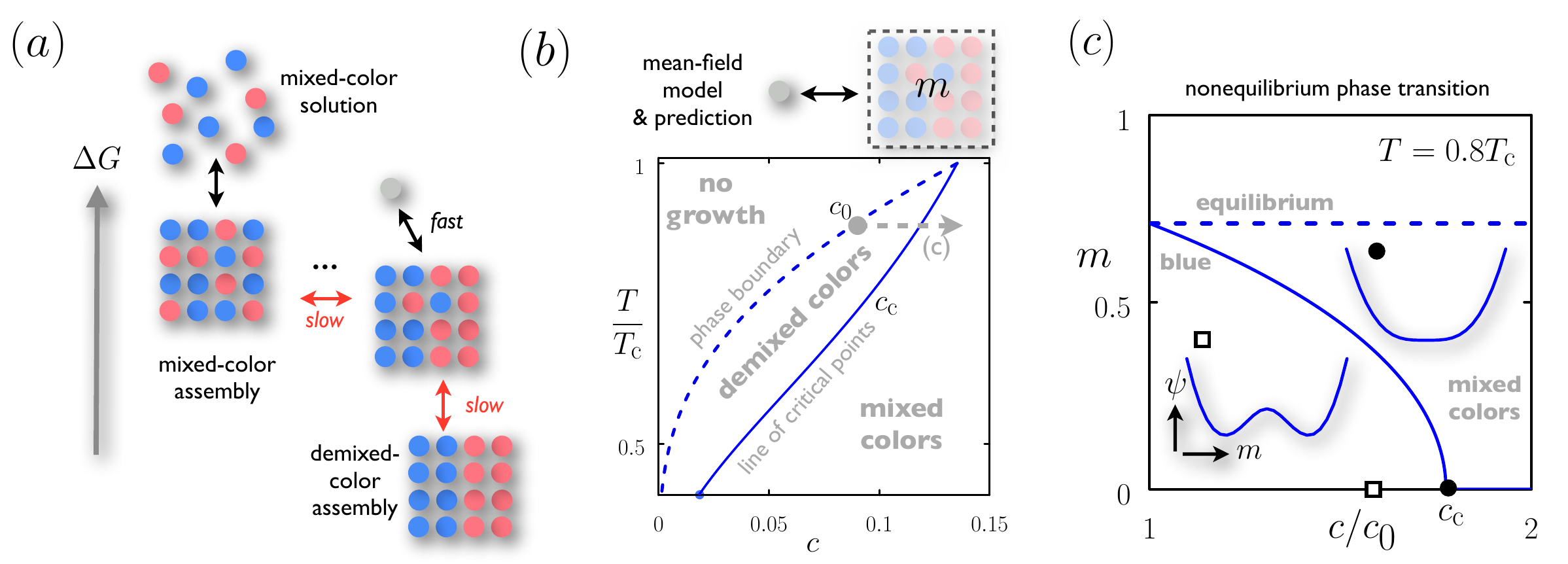}
\caption{\label{fig1} {\em Mean-field theory predicts a nonequilibrium phase transition within a growing two-component structure}. (a) The conceptual growth process modeled in this paper: Particles possess a free energetic impetus ($\Delta G$) to assemble in a phase-separated fashion, but the sluggish dynamics of component type rearrangement (red arrows) can thwart such equilibration. See text for description of this process. (b) A mean-field model of this process (sketch) predicts the displayed dynamic phase diagram in the temperature-concentration plane (parameters: $\ed= \es/3$). Below the equilibrium critical temperature $T_{\rm c}$, the structure in equilibrium (along the dashed line $c=c_0$) is demixed. Increasing concentration at fixed temperature (dashed arrow in panel (b)) causes the assembly to grow and become more mixed, and eventually to undergo a dynamic phase transition to a mixed-color structure (panel (c)). This prediction is borne out by simulations (Figs. 2 \& 3). Square and circle symbols on panel (c) link the potential $\psi$ (shown inset) to the concentrations at which it was calculated.}
\end{figure*}

Predicting the outcome of such `far-from-equilibrium' self-assembly is not generally possible. For the reasons just described, we cannot do so within the framework of equilibrium statistical mechanics, i.e. by calculating and inspecting the appropriate free energy landscape. Instead, we need explicit information about the dynamics undergone by components, together with a predictive theory that works `far' from equilibrium. Despite much progress in this direction\c{kremer1978multi,stauffer1976kinetic,PhysRevB.27.7372,schmelzer2004nucleation,schmelzer2000reconciling,scarlett2010computational,kim2008probing,scarlett2011mechanistic,sanz2007evidence,peters2009competing}, such a theory does not exist. Here we present evidence suggesting that concepts of nonequilibrium statistical physics may provide a route to a predictive theory of far-from-equilibrium self-assembly. We study two-component lattice-based assemblies made of red and blue particles, essentially growing versions of the Ising lattice gas. We find -- using dynamic mean-field theory and simulation -- that these assemblies undergo, as a function of growth rate, a nonequilibrium phase transition between mixed and demixed arrangements of component types.  If component interactions are chosen to mimic the equilibrium ferromagnetic Ising model {\em below} its critical temperature, then the equilibrium structure of an assembly is a demixed one of phase-separated domains of red and blue. Assemblies grown at finite rate possess patterns more mixed than the equilibrium one, and at a certain growth rate one encounters a dynamic critical point, beyond which component types mix within the assembly in a manner similar to that of the (anisotropic) equilibrium Ising model {\em above} its critical temperature. Our results suggest a way to generate qualitatively defined two-component structures using, for instance, DNA-linked colloids\c{kim2008probing,scarlett2010computational}. Such experiments would serve as a test of the predictions of this paper.

We shall model the conceptual two-component growth process (ignoring nucleation) sketched in \f{fig1}(a). We assume 1) that red and blue particles are present in solution in equal number, and are randomly mixed; 2) that red and blue particles possess energetic interactions that allow them to associate but encourage them to phase separate; 3) that red and blue particles may bind to a growing structure (called an `assembly') and unbind from it, 4) thereby modifying the composition (red and blue patterns) of that assembly; and 5) that the mobility of red and blue particles {\em within} the assembly is low, because it it solid. We first describe this process within mean-field theory, accounting for each assumption in a rough manner. We then introduce a lattice-based simulation model of this growth process, which resolves fluctuations and microscopic detail omitted by the mean-field theory. 

In \f{fig1}(b) we sketch a mean-field model of this growth process. Only two types of microscopic process are permitted: a particle may bind to an assembly, or unbind from it (Assumption 3). We shall resolve only the rates for these processes, not the individual microscopic events. The assembly itself is described only by a parameter $m \in [-1,1]$, similar to an Ising magnetization: when $m=-1$ the assembly is all red; when $m=1$ the assembly is all blue; and when $m=0$ the assembly is of mixed color. We further assume that red and blue particles add to the assembly with equal rates $c/2$ (Assumption 1), $c$ being proportional to the solution concentration of particles. We assume that particles leave the assembly with a rate proportional to the abundance of the relevant color within the assembly, multiplied by the Boltzmann weight of the energy of particle removal from the assembly. The latter is computed as follows. To allow us to account for Assumption 2, we set the red-red and blue-blue interactions to $-\es$ (`same'), and the red-blue interaction to $-\ed$ (`different'). We assume that each particle within the assembly makes $z$ energetic bonds, where $z$ is constant (i.e. we do not distinguish between the bulk and surface of the assembly).  We consider bonds to connect to a red particle with probability $p_{\rm R} = (1-m)/2$, and to a blue particle with probability $p_{\rm B} = 1-p_{\rm R}$. The energy of interaction per bond between a particle of color $\alpha$ ($\alpha=$ B or R) and the assembly is then $\epsilon_\alpha = -\es p_\alpha - \ed (1-p_\alpha)$. Putting these pieces together, we have that the net rates of particle attachments are
\beq
\label{rate}
\Gamma_\alpha = c/2- p_\alpha \exp \left( \beta z  \epsilon_\alpha  \right),
\eeq
where $\alpha=$ B or R, and $\beta \equiv 1/\left(\kt\right)$. Finally, we solve for the composition $m$ of the assembly by setting $\Gamma_{\rm R}/\Gamma_{\rm B} = p_{\rm R}/p_{\rm B}$, which is the self-consistent requirement that the ratio of growth rates of red and blue particles be equal to the relative abundance of red and blue particles in the assembly. This requirement accounts for Assumption 4. Furthermore, because {\em only} binding-unbinding processes determine the composition of the assembly, we also account for Assumption 5: specifically, the processes labeled by the red arrows in \f{fig1}(a) are assumed to happen with zero rate in this mean-field theory.
\begin{figure}[t!]
\includemovie[poster=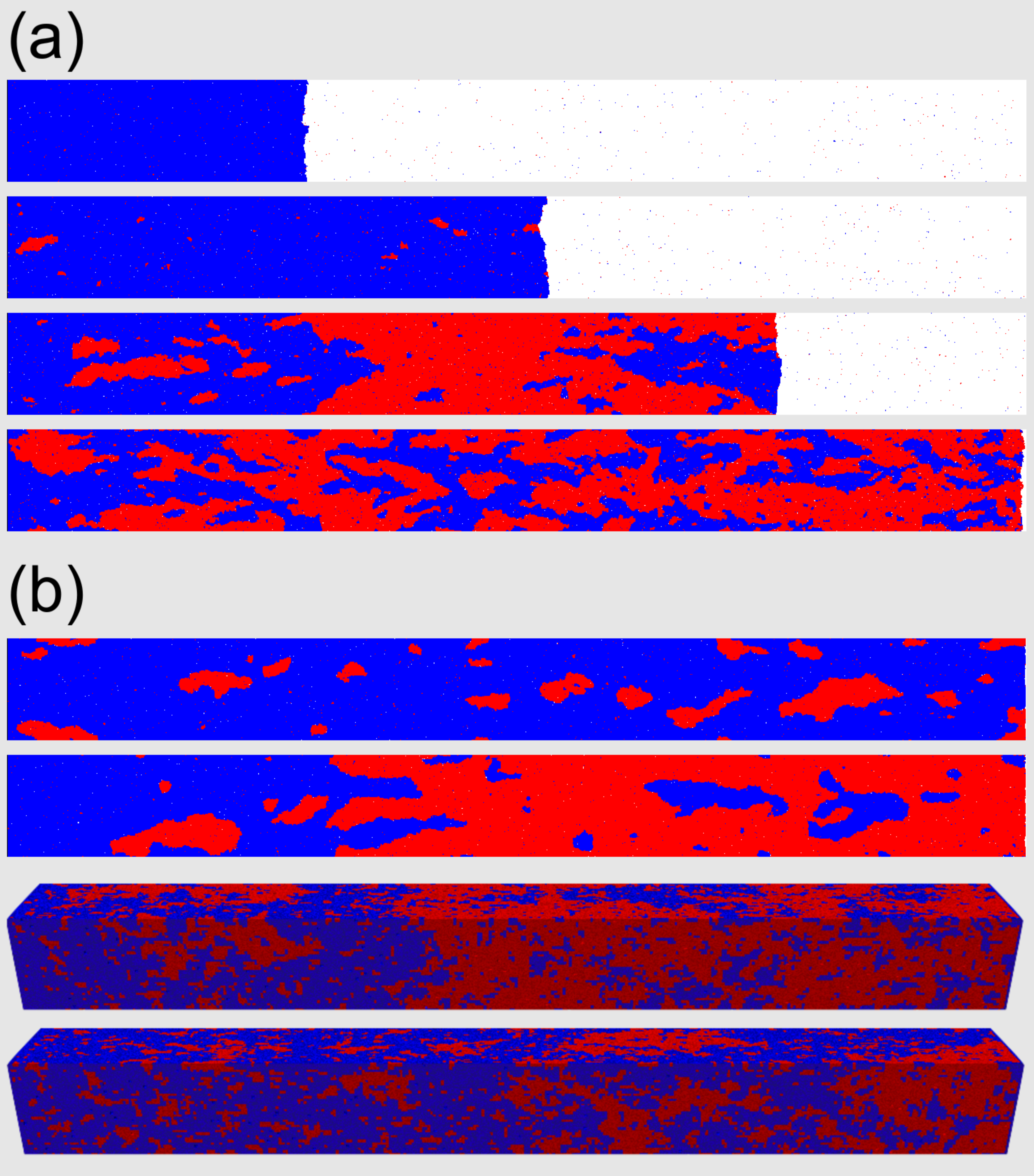, mouse]{\linewidth}{1.137\linewidth}{fig2.mp4}
\caption{\label{fig2} {\em Snapshots and movies illustrating a nonequilibrium phase transition in 2D and 3D simulations of self-assembly}. [If viewing in Adobe Reader, click figure to animate] (a) Two-component growth in two dimensions. Assemblies generated at increasing rates of growth (top to bottom) reveal the existence of a dynamical phase transition between near-equilibrium demixed structures (top), and mixed-color structures (bottom). At a particular growth rate (second bottom), color structures appear critical. Snapshots are taken at a time corresponding to the completion of growth in the bottom panel. (b) Representative snapshots of growth patterns at dynamic criticality in two- and three dimensions, illustrating the large run-to-run variations in color structure.}
\end{figure}

It is straightforward to verify that \eqq{rate}, in the equilibrium (zero growth rate) limit, describes the usual Ising-like equilibrium red-blue demixing phase transition when $\es-\ed$ is large enough\c{chandler} (Fig. S1). This phase transition happens along the line labeled $c_0$ in \f{fig1}(b), on which the assembly neither grows nor shrinks. If we start on this line and increase concentration $c$ in order to make the assembly grow, then one can show (\href{http://nanotheory.lbl.gov/people/criticality_paper/criticality_supp.pdf}{see SI}) that the composition $m$ of the assembly is described by a particular Landau theory\c{binney1992theory},
\bea
\label{eq_psi}
\psi(m) = \frac{1}{2}\left( 1-\frac{\Delta}{\nu} \right) m^2 
+ \frac{\Delta}{24} \left( \frac{6-\nu^2}{\nu^3}\right) m^4 + \cdots,
\eea 
in which $\nu \equiv c \, \e^{\Sigma}$ is a rescaled concentration, $\Delta \equiv \beta z \left( \es-\ed \right)/2$, and $\Sigma \equiv \beta z \left( \es+\ed \right)/2$. This description makes explicit the connection between this growth problem and the field of nonequilibrium critical phenomena\c{binney1992theory}. The minima of the potential $\psi(m)$ describe qualitatively the nonequilibrium color patterns generated during growth of the assembly. Inspection of the coefficients of \eqq{eq_psi} reveals the behavior shown in \f{fig1}(b). When assembly color patterns are demixed in equilibrium, there exists a line of {\em nonequilibrium} continuous phase transitions at a critical concentration given by $\nu = \Delta$, i.e. $c_{\rm c} = \Delta \e^{-\Sigma}$ (provided that $\nu<  \sqrt{6}$). \f{fig1}(c) shows that $\psi$ changes from having one well to having two wells at this dynamic critical point. 

The prediction of mean-field theory is therefore as follows: if we are below the equilibrium critical temperature, then the two-component structure {\em in equilibrium} is a demixed one of red or blue. If we supersaturate the assembly in order to make it grow, then it becomes more mixed, and at a critical growth rate undergoes a dynamic phase transition to a mixed-color structure. If we are not too far below the equilibrium critical temperature then this dynamic phase transition is a continuous one, with an associated dynamic critical point.

Although mean-field theory is not quantitatively accurate, this qualitative prediction is confirmed by our simulations of two-component self-assembly. We again model the conceptual growth process sketched in \f{fig1}(a), this time in microscopic detail. We considered square and cubic lattices in 2D and 3D, respectively, within rectangular or cuboidal simulation boxes (periodic boundaries were imposed across the short axes only). Lattice sites could be white (unoccupied), or occupied by a red or blue particle. We imposed pairwise nearest-neighbor interactions between particles, $-\es$ for like-color interactions, and $-\ed$ for unlike-color interactions. Each particle is disfavored energetically by a chemical potential term $\mu'>0$. We chose $\mu'$ small enough that equilibrium configurations corresponded to almost fully-occupied lattices, whose equilibrium red-blue patterns are those of the 2D or 3D Ising model with ferromagnetic coupling $J=(\es-\ed)/2$. We chose $\beta J=0.75$ (2D) and $\beta J=0.275$ (3D) so that equilibrium color patterns were demixed (Assumption 2).

To model a growth dynamics, we used a simulation protocol that satisfies detailed balance but makes no assumptions about relative rates of growth and structural relaxation\c{drossel1997model}. We chose at random a lattice site. If white, we attempted to make it red or blue, with equal likelihood (Assumption 1), and accepted this proposition with probability $\min\left( 1, {\rm e}^{-\beta \Delta E-\beta \mu} \right)$. Here $\Delta E$ is the interaction energy felt by the newly-placed particle, and $\mu = -\kt \ln 2 + \mu'$ (see SI). If the randomly-chosen lattice site was instead blue or red, we made it white with probability $\min\left( 1, {\rm e}^{-\beta \Delta E+\beta \mu} \right)$, where $\Delta E$ is the energy change upon removing the particle from the simulation box. Assumptions 3 and 4 are naturally accounted for by these microscopic processes. Finally, to account for Assumption 5, we imposed a kinetic constraint that prevents any change of state of a lattice site having exactly $2d$ occupied neighbors. This constraint prevents relaxation of color patterns within the bulk of an assembly, except in the neighborhood of a vacancy (the processes described by the red arrows in \f{fig1} are then slow, but do happen; see SI). The constraint respects detailed balance, and so has no effect on the thermodynamics of the model. 
\begin{figure}[t!]
\centering
\includegraphics[width=\linewidth]{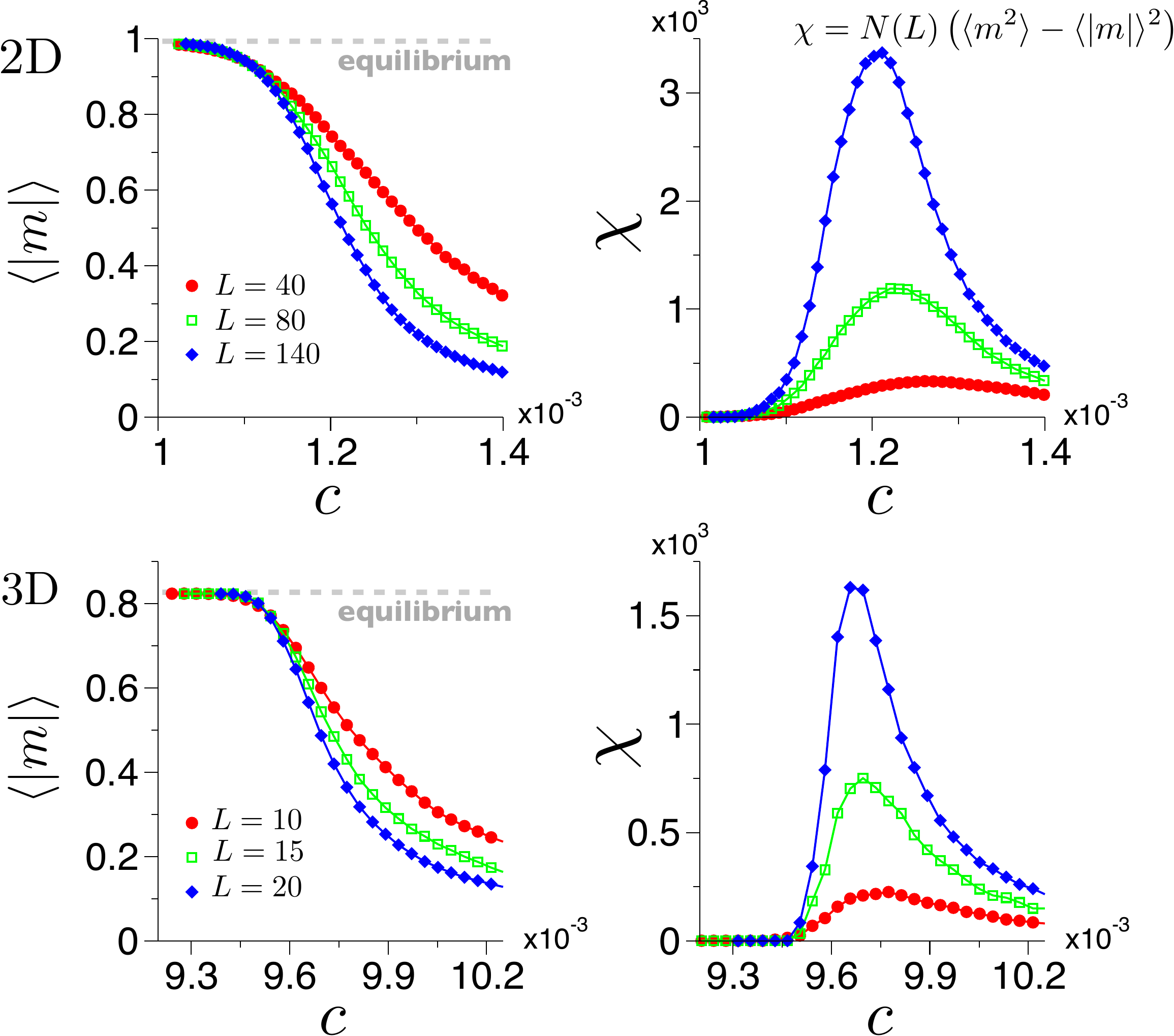}
\caption{\label{fig3} {\em Simulations confirm the existence of a nonequilibrium critical point at a particular growth rate}. We plot color order parameter $m$ (the density of blue particles minus the density of red particles within the assembly) and color fluctuations $\chi$ as a function of basic rate $c=\exp(-\beta \mu)$. Here $N(L)$ is the number of particles in the sampling volume (\href{http://nanotheory.lbl.gov/people/criticality_paper/criticality_supp.pdf}{see SI} for details). The peak in $\chi$ indicates large fluctuations characteristic of criticality.}
\end{figure}

The parameter $c \equiv \exp(-\beta \mu)$ controls the basic rate of growth of an assembly (\href{http://nanotheory.lbl.gov/people/criticality_paper/criticality_supp.pdf}{Fig. S2}). We began each growth simulation from a white box. By making one wall of the simulation box short axis sticky, and by choosing $c$ not too large, we could cause the growth of an assembly in the direction of the long axis of the box, without having clusters nucleate elsewhere. Moreover, by making the wall sticky for only blue particles, we could induce a color symmetry-breaking that made establishment of the steady-state color structure of the assembly relatively rapid. 

In Figs. \ref{fig2} and \ref{fig3} we show that simulations confirm the prediction of mean-field theory. At low rates of growth, assemblies are almost blue, because of the way we intentionally break symmetry, with only a few red cluster defects: see \f{fig2}(a). Given the large lengthscales involved in equilibrium phase separation, and our relatively small system sizes, we lack the precision to determine if structures grown in this region are truly the equilibrium one. In terms of the parameter $m$, they appear to be close to it: see \f{fig3}, left panels, and \href{http://nanotheory.lbl.gov/people/criticality_paper/criticality_supp.pdf}{Fig. S2}. Note, however, that equilibrium structures are effectively one-component ones (because we are below the critical temperature and we work in the grand-canonical ensemble), and so slow growth in this regime is not a viable way of making interesting two-component patterns. At larger growth rates the assemblies harbor larger red clusters, and at a particular growth rate assemblies appear critical, in the sense that from run to run their patterns exhibit large color fluctuations: see \f{fig2}(b) and \f{fig3}, right panels. At all rates of growth the thermodynamically stable assembly is mostly blue, and so the observed mixing reflects the existence of a nonequilibrium phase transition\c{park2012surface}. 

In previous work we demonstrated that simple nonequilibrium arguments allow one to predict the patterns made within a growing two-component 1D fiber\c{whitelam2012self}. Here we considered self-assembly in 2D and 3D, which is more directly relevant to experiment. In all cases we find that simple nonequilibrium arguments can make progress where equilibrium ideas have little to say. The case $d>1$ is intrinsically richer, however, displaying nonequilibrium phase transitions not seen in $d=1$ (similar to well-known equilibrium behaviors in different dimensions\c{binney1992theory}). Additionally, in $d=1$ there exists an exact mapping between structures in and out of equilibrium, allowing one to describe patterns generated at finite rate of growth in terms of an equilibrium system with `renormalized' coupling constants, but the same cannot be true here: structures seen in snapshots (\f{fig2}) are visibly anisotropic\c{drossel1997model}, reflecting a memory of the assembly's growth direction. Therefore, if there exists an equivalent equilibrium system~\cite{grinstein1985statistical,whitelam2012self}, it has an anisotropic energy function. Exploring this equivalence is a subject for future work.

We have shown that patterns formed within growing two-component assemblies undergo, as a function of growth rate, a nonequilibrium phase transition in their component type arrangements. Qualitatively similar transitions are seen in certain irreversible cellular automata upon variation of interaction parameters\c{ausloos1993magnetic,candia2008magnetic}, pointing to a possible connection, away from equilibrium, between irreversible automata and the microscopically reversible simulation protocols used to model self-assembly. Experimentally, the predictions of our paper might be tested using two-component DNA-linked colloids\c{kim2008probing,scarlett2010computational}. Our results suggest that one can assemble (Fig. S3) complex mesostructures\c{bahatyrova2004native} by exploiting the intrinsically far-from-equilibrium nature of multicomponent self-assembly, even under conditions that give rise to mundane {\em equilibrium} patterns.

{\em Achnowledgements.} We thank Rob Jack for discussions. This work was done as part of a User project at the Molecular Foundry at Lawrence Berkeley National Lab, and was supported by the Office of Science, Office of Basic Energy Sciences, of the U.S. Department of Energy under Contract No. DE-AC02--05CH11231. L.O.H. was supported by the Center for Nanoscale Control of Geologic CO$_2$, a U.S. D.O.E. Energy Frontier Research Center, under Contract No. DE-AC02--05CH11231. J.D.S. acknowledges support from KSU startup funds. This research used resources of the National Energy Research Scientific Computing Center, which is supported by the Office of Science of the U.S. Department of Energy under Contract No. DE-AC02-05CH11231.


\end{document}